\documentclass[aps,preprint,prd,showpacs,showkeys,nofootinbib,superscriptaddress]{revtex4-1}
\usepackage{amsfonts}
\usepackage{mathrsfs}
\usepackage{graphicx}
\usepackage{amsmath}
\usepackage{amssymb}
\usepackage{bm}
\usepackage{multirow}
\usepackage{epsfig}
\usepackage{subfig}
\usepackage{hyperref}
\usepackage{ulem}
\usepackage{color}

\hypersetup{hidelinks}

\begin{document}
\parskip +7pt

\title{Where to find $X(17)$?}

\author{Jun Jiang}
\email{jiangjun87@sdu.edu.cn}
\affiliation{School of Physics, Shandong University, Jinan, Shandong 250100, China}

\author{Cong-Feng Qiao}
\email{qiaocf@ucas.ac.cn.}
\affiliation{School of Physical Sciences, University of Chinese Academy of Sciences, Beijing 100049, China}

\author{Yu-Han Zhao}
\email{zhaoyh112358@163.com}
\affiliation{Public Courses Teaching Department, Shaanxi Polytechnic University, Xianyang, Shaanxi 712000, China}

\date{\today}

\begin{abstract}

The Atomki group puts forward the hypothesis of an $X(17)$ particle to explain its anomaly observation in ${^{8}\text{Be}^{*}} \to {^{8}\text{Be}} e^+e^-$ transition.
Utilizing experimental results from the Atomki experiments, measurements of the electron's anomalous magnetic moment (AMM), the parity-violating M{\o}ller scattering experiment, the NA64 experiment, the KLOE-2 experiment, and the PADME experiment, we derive constraints on the $Xee$ coupling of the hypothetical $X(17)$ boson to electrons. 
A joint $\chi^2$ analysis of the electron's AMM (Cs and Rb measurements) and the parity-violating M{\o}ller scattering yields two mutually exclusive solutions. 
The Cs and Rb AMM constraints restrict the combination $(\varepsilon_e^v)^2-5(\varepsilon_e^a)^2$ with opposite signs. Cs+M{\o}ller combination favours a dominant axial-vector coupling $|\varepsilon_e^a|\approx5.3\times10^{-4}\gg|\varepsilon_e^v|$, whereas Rb+M{\o}ller favours a dominant vector coupling $|\varepsilon_e^v|\approx6.9\times10^{-4}\gg|\varepsilon_e^a|$. 
The M{\o}ller bound suppresses the subdominant coupling for both solutions. 
Incorporating NA64 and KLOE-2 limits on $(\varepsilon_e^v)^2+(\varepsilon_e^a)^2$ introduces mild tension for the Cs+M{\o}ller solution, while the Rb+M{\o}ller solution remains nearly unchanged.

\vspace {7mm} 

\end{abstract}
\maketitle

\section{Introduction}
\label{sec:Introduction}

The Standard Model (SM) explains the fundamental interactions between elementary particles almost perfectly.
However, phenomena such as the matter-antimatter asymmetry, dark matter, hierarchy problem, strong charge-parity problem, and the nature of the Higgs boson, along with flavor anomalies, indicate that New Physics (NP) must exist and is concealed within the anomalies beyond SM.
These mysteries and experimental anomalies motivate physicists to extend the SM and explore new forces and mechanisms.

In 2016, Krasznahorkay {\it et al.} from Atomki reported the observation of an unexpected bump in the electron-positron angular correlations during the transition of the excited ${^{8}\text{Be}^{*}}$ nucleus to its ground state \cite{Krasznahorkay:2015iga,Krasznahorkay:2018snd}. 
They proposed that this bump arose from the production and subsequent decay of a neutral isoscalar particle with a mass of approximately 17 MeV into an electron-positron pair, termed $X(17)$.
In 2019, the Atomki group investigated the energy-sum and angular correlation spectra of electron-positron pairs emitted during the transition of excited ${^4\text{He}^*}$ to its ground state, and observed a peak that could also be explained by the previously proposed hypothetical $X(17)$ particle \cite{Krasznahorkay:2019lyl,Krasznahorkay:2019lgi,Krasznahorkay:2021joi}.
In 2022, they discovered a similar anomaly in the transition of excited $^{12}\text{C}^*$ to its ground state, which could be again well accounted for by the existence of the hypothetical $X(17)$ \cite{Krasznahorkay:2022pxs}.
In 2023, they further confirmed the existence of $X(17)$ in an experiment investigating the Giant Dipole Resonance of ${^{8}\text{Be}^{*}}$ \cite{Krasznahorkay:2023sax,Krasznahorky:2024adr}.

Several other experiments have attempted to replicate the ${^{8}\text{Be}^{*}}$ anomaly by Atomki or to search for $X(17)$ in new experimental setups. 
The VNU-UoS experiment reproduced the anomaly in ${^{8}\text{Be}^{*}}$ decay and confirmed the $X(17)$ resonance with a significance exceeding $4\sigma$ \cite{Arias-Aragon:2025wdt}. 
In contrast, the MEG-II experiment also repeated the ${^{8}\text{Be}^{*}}$ measurement but found no significant signal; instead, it derived an upper limit on the $X(17)$ production branching ratio relative to photon emission \cite{MEGII:2024urz}.
The PADME experiment at the Frascati DA$\Phi$NE LINAC searched for $X(17)$ using a positron beam impinging on a fixed target \cite{PADME:2025dla}. 
Within a beam energy range of 262 MeV to 296 MeV (corresponding to center-of-mass energies of $16.4 < \sqrt{s} < 17.4$ MeV), the production rates of the Bhabha scattering process $e^+e^- \to e^+e^-$ and the $e^+e^- \to \gamma\gamma$ process were measured. 
A significant deviation was observed at $\sqrt{s} = 16.90$ MeV, with a local significance of $2.5\sigma$ and a global significance of $(1.77 \pm 0.15)\sigma$. 
More experiments are on the way to search for $X(17)$ signals \cite{Alves:2023ree}.

The Atomki anomalies have attracted significant interests from theoretical physicists in both nuclear physics and particle physics. 
Feng {\it et al.} proposed a protophobic fifth-force interpretation for the observed anomaly in ${^8\text{Be}}$ nuclear transitions \cite{Feng:2016jff,Feng:2016ysn,Feng:2020mbt}, which brought the Atomki anomaly to the forefront of the NP community. 
The most puzzling aspect of the Atomki anomalies is the extremely low energy scale, which is clearly outside the realm of Quantum Chromodynamics. 
A very light and very weakly coupled, generalized version of a dark photon and now a possible $X(17)$ candidate, starting from 1980, was proposed by Fayet \cite{Fayet:1980ad,Fayet:1980rr,Fayet:2020bmb}.
Wong argued that the Quantum Electrodynamics (QED) mesons (quark-antiquark pairs interacting solely via QED) can explain the Atomki anomaly \cite{Wong:2020hjc,Wong:2021blz}. 
Zhang {\it et al.} interpreted the anomaly using the nuclear transition form factor \cite{Zhang:2017zap}. 
Subsequently, they found that $X(17)$ production is dominated by direct transitions induced by transverse and longitudinal electric dipoles as well as charge dipoles, without proceeding through any nuclear resonance with a smooth energy dependence that occurs for all proton beam energies above threshold \cite{Zhang:2020ukq}. 
Hayes {\it et al.} reexamined the angular correlations in the $e^+e^-$ decay of ${^{8}\text{Be}^{*}}$ and found no evidence for $X(17)$ \cite{Hayes:2021hin}. 
They found that the existence of a ``bump'' in the measured angular distribution was strongly dependent on the assumed ratio of M1 transition to E1 transition, which was a strong function of energy, while it was assumed to be a constant in the Atomki experiments.
To date, numerous NP explanations have been shown to be invalid, while several proposed solutions remain viable, see Ref. \cite{Alves:2023ree} for a brief review.

Physicists have further attempted to derive constraints on the couplings of $X(17)$ to electrons and nucleons.
In Refs. \cite{Feng:2016jff,Feng:2016ysn}, Feng \textit{et al.} discuss a protophobic vector gauge boson interpretation of the Atomki anomaly. For a vector $X(17)$, they constrain the electron coupling strength $\varepsilon^v_e$ to the range $2\times 10^{-4} \lesssim |\varepsilon^v_e| \lesssim 1.4 \times 10^{-3}$, while the neutron coupling strength is constrained to $|\varepsilon^v_n| \lesssim 2.5 \times 10^{-2}$.
Using a combined analysis of data samples collected in 2017 and 2018, the NA64 collaboration reports that the parameter region $1.2\times10^{-4}<|\varepsilon^v_e|<6.8\times10^{-4}$ can be excluded at the 90\%  confidence level (C.L.) \cite{Banerjee:2019hmi}. 
The PADME experiment reports an excess at $\sqrt{s} = 16.9$ MeV corresponding a significance of approximately $2\sigma$, and obtains a coupling strength of $|\varepsilon^v_{e}| \approx 1.85 \times 10^{-3}$ at the 90\%  C.L. \cite{PADME:2025dla}.\footnote{The definition of the coupling strength in the PADME experiment \cite{PADME:2025dla} differs by an electric charge $e$ from that in Refs. \cite{Feng:2016jff,Feng:2016ysn} by Feng \textit{et al.}, that in the NA64 experiment \cite{Banerjee:2019hmi}, and our definition in Eq. \eqref{eq:lagrangian} below. We rescale the relevant values consistently throughout the paper to enable direct comparisons.} 
In Ref. \cite{DiLuzio:2025ojt}, Luzio \textit{et al.} claim that the excess observed by PADME for the vector coupling is already in tension with constraints from two sources: the electron's anomalous magnetic moment (AMM), and the non-observation of exotic decays $\pi^+ \to e^+\nu X(17)$ and $\mu^+ \to e^+ \bar{\nu}_\mu \nu_e X(17)$ at the SINDRUM experiment.
In Ref. \cite{Krasnikov:2019dgh}, Krasnikov points out that the NA64 bounds on the $Xee$ coupling, combined with the latest values of the electron's AMM \cite{Parker:2018vye}, exclude pure vector or axial-vector couplings of the $X(17)$ boson to electrons at the 90\% C.L.; however, mixed ``vector $\pm$ axial-vector'' ($V \pm A$) couplings remain viable.
For the $V \pm A$ coupling to electrons, Barducci and Toni discuss constraints on the parameter $\sqrt{(\varepsilon^v_e)^2 + (\varepsilon^v_a)^2}$ extracted from multiple experiments in the Appendix F of Ref. \cite{Barducci:2022lqd}.
We will compare our results with theirs in Section \ref{sec:constrain_model}.\footnote{The definition of coupling strength in Ref. \cite{Barducci:2022lqd} by Barducci and Toni also differs by an electric cahrge $e$ from ours.} 
Very recently, Fieg \textit{et al.} considered the chiral couplings ({\it i.e., $V \pm A$}) of $X(17)$ to quarks and found that the 99\% C.L. parameter region is in tension with experimental constraints, where the tension is driven by the $^{12}\text{C}$ experiment by Atomki \cite{Fieg:2026zkg}.

In the subsequent Section \ref{sec:constrain_model}, we will derive constraints on the coupling parameters of $X(17)$ to electrons using data from the Atomki experiments, the electron's AMM, the parity-violating M{\o}ller scattering experiment, beam dump experiments, the KLOE-2 experiment, and the PADME experiment.
We will focus on discussion of the parameter space for the  $V \pm A$ model describing the couplings of $X(17)$ to electrons. 
Section \ref{sec:sum} is devoted to a summary of our findings and an outlook on future prospects.

\section{Constraints on $X(17)$ couplings to electrons}
\label{sec:constrain_model}

In Subsection \ref{subsec:VAmodel}, we discuss the models of $X(17)$ couplings to electrons: pseudoscalar, scalar, vector, axial-vector, and particularly vector $\pm$ axial-vector ($V \pm A$).
We then derive constraints on the parameter space of the $V \pm A$ model from the Atomki experiments, the measurement of electron's anomalous magnetic moment (AMM), the parity-violating M{\o}ller scattering experiment, beam dump experiments, the KLOE-2 experiment, and the PADME experiment in Subsection \ref{subsec:constrainee}.
The conclusions are drawn in Subsection \ref{subsec:survivalee}.

\subsection{``Vector $\pm$ Axial-vector'' Model}
\label{subsec:VAmodel}

The Atomki experiments have indicated that the $X(17)$ particle couples to the electrons definitely. 
Then what would be the interaction models for the $Xee$ coupling: pseudoscalr, scalar, vector, axial-vector, or vector $\pm$ axial-vector ($V \pm A$)?
The scalar $X(17)$ hypothesis has been excluded by parity conservation in the Atomki ${^8\text{Be}}$ experiment, while a pseudoscalar state can only explain the ${^8\text{Be}}$ and ${^4\text{He}}$ anomalies, not the $^{12}\text{C}$ one \cite{Barducci:2022lqd}.
In Ref. \cite{Krasnikov:2019dgh}, Krasnikov argues that the combination of results from the latest NA64 experiment \cite{Banerjee:2019hmi} and the recent value of the electron's anomalous magnetic moment (AMM) \cite{Parker:2018vye} excludes pure vector and pure axial-vector couplings of the $X(17)$ boson to electrons at the 90\% C.L. However, the model with ``$V \pm A$'' interaction remains viable and can further account for both the electron and muon AMMs.\footnote{The latest analysis of the muon's AMM indicates that there is no tension between the Standard Model (SM) prediction \cite{Aliberti:2025beg} and experimental results \cite{Muong-2:2025xyk} at the current level of precision.} 
We will discuss the constraints on these interaction models specifically under available experiments in Subsection \ref{subsec:constrainee}.

In our previous work \cite{Jiang:2018uhs,Jiang:2018jqp}, we adopted the ``$V-A$'' model to describe the interaction between the $X(17)$ boson and fermions, and made a rough assumption of equal vector and axial-vector coupling strengths in the numerical calculations. 
In the present manuscript, we extend this hypothesis to the general ``$V \pm A$'' model with the general Lagrangian,
\begin{equation}
    \mathcal{L}_X=-\frac{1}{4}X_{\mu\nu}X^{\mu\nu} + \frac{1}{2}m_X^2X_{\mu}X^{\mu}-\sum_f e\bar{f}\gamma_\mu (\varepsilon^v_f-\varepsilon^a_f\gamma_5) f X^\mu,
\label{eq:lagrangian}
\end{equation}
where the hypothetical $X(17)$ with mass $m_X$ has the field strength $X_{\mu\nu} \equiv \partial_{\mu} X_{\nu}-\partial_{\nu} X_{\mu}$. 
The coupling strengths, denoted $\varepsilon^{v/a}_f$, are given in units of the electronic charge $e$. 
Here, the coupling parameters $\varepsilon^{v/a}_f$ can take either positive or negative values, implying that the interaction exhibits ``$V \pm A$'' coupling to the fermion $f$.
In the present manuscript, we focus exclusively on the coupling $Xee$ of $X(17)$ boson to electrons.

\subsection{Constraints on $Xee$ coupling}
\label{subsec:constrainee}

\subsubsection{Atomki Experiments}

In 2016, Krasznahorkay {\it et al.} reported the production of a neutral particle $X$, which subsequently decays to $e^+e^-$ via internal pair creation, along with its mass and the ratio of decay widths \cite{Krasznahorkay:2015iga},
\begin{align}
&m_{X}= 16.7 \pm 0.35(\text {stat.}) \pm 0.5(\text {sys.}) \text { MeV}, \\
&\frac{\Gamma\left( ^{8}\text{Be}^{*} \rightarrow {^{8}\text{Be}} X\right)\operatorname{Br}\left(X \rightarrow e^{+} e^{-}\right)}{\Gamma\left( ^{8}\text{Be}^{*} \rightarrow {^{8}\text{Be}} \gamma\right)} =5.8 \times 10^{-6}. \label{eq:ratiobe}
\end{align}
Later, they updated the measurement of its mass to $m_X=17.01\pm0.16$ MeV and the branching fraction ratio in Eq. (\ref{eq:ratiobe}) to $(6\pm1)\times10^{-6}$ \cite{Krasznahorkay:2018snd,Krasznahorkay:2019lgi}.
Using $\Gamma\left(^{8}\text{Be}^{*} \rightarrow{^{8}\text{Be}} \gamma\right)=1.9\pm0.4$ eV \cite{Tilley:2004zz} and further assuming $\operatorname{Br}\left(X \rightarrow e^{+} e^{-}\right) \approx 1$, we obtain $\Gamma\left(^{8}\text{Be}^{*} \rightarrow{^{8}\text{Be}} X\right)=(1.1\pm0.31)\times10^{-5}$ eV.
The radiative transition of $^8\text{Be}^*$ ($J^P=1^+$) to the ground state $^8\text{Be}$ ($J^P=0^+$) implies that $X(17)$ can be a $J^P=1^+$ boson (orbital angular momentum $L=0$), or $J^P=1^-$ boson ($L=1$), or a $J^P=0^-$ boson ($L=1$).
Very recently, combined results from the PADME experiment with Atomki nuclear physics results, Arias-Arag$\Acute{\mathrm{o}}$n {\it et al.} derive a value of $m_X = 16.88 \pm 0.05$ MeV for the $X(17)$ mass \cite{Arias-Aragon:2025wdt}, which will be adopted throughout this paper.

In the rest frame of the ${^{8}\text{Be}^{*}} \rightarrow {^{8}\text{Be}} X$ decay\footnote{In the laboratory frame, the produced $^8\text{Be}^*$ state moves nonrelativistically with a velocity of $0.006c$, where $c$ is the speed of light. Thus, assuming the $^8\text{Be}^*$ rest frame is appropriate.}, the $X(17)$ boson is produced with a velocity of $v\approx0.37c$. One can then estimate the decay length for the $X \to e^+e^-$ process as
\begin{equation}
    L = \frac{\hbar \times c}{\Gamma(X \rightarrow e^{+} e^{-})}\times\frac{v}{\sqrt{1-v^2}},
\end{equation}
where $\hbar$ is the reduced Planck constant. Here, we assume that the $X \rightarrow e^+e^-$ decay is saturated. Within the ``$V \pm A$'' model, the decay width of the $X$ boson to electron-positron pairs can be formulated as follows, up to an error of $\mathcal{O}\left(\frac{m_e^2}{m_X^2}\right)\approx0.1\%$,
\begin{equation}
    \Gamma(X\rightarrow e^+e^-)\approx \frac{\alpha m_X}{3}\big((\varepsilon^v_e)^2+(\varepsilon^a_e)^2\big),
\label{eq:decaywidth_approx}
\end{equation}
where $\alpha$ is the fine structure constant. This implies that the vector and axial-vector couplings make identical contributions aside from differences in their coupling strengths. This approximate expression in Eq. \eqref{eq:decaywidth_approx} is sufficiently accurate, and its distinction from the exact result is negligible, which we have illustrated in our previous work \cite{Jiang:2018uhs,Jiang:2018jqp}.

According to the setup of the Atomki experiments \cite{Gulyas:2015mia}, the travel length of the $X(17)$ boson must be less than approximately 5 cm. Numerically, imposing the constraint $L \leq 5$ cm on the decay length, the $^8\text{Be}$ anomaly experiment implies $(\varepsilon^v_e)^2 + (\varepsilon^a_e)^2 \gtrsim 3.8 \times 10^{-11}$.
If we assume that the $X$ boson decays shortly after its production and further impose the stricter constraint $L \leq 1$ cm, we obtain
\begin{equation}
    (\varepsilon^v_e)^2 + (\varepsilon^a_e)^2 \gtrsim 1.9 \times 10^{-10}.
    \label{constrainee1}
\end{equation}
For the pure vector coupling case (setting $\varepsilon^a_e = 0$), we derive $|\varepsilon^v_e| \gtrsim 1.4 \times 10^{-5}$, which is consistent with the estimate in Ref. \cite{Feng:2016jff} when adopting $m_X = 17$ MeV.

In 2019, Krasznahorkay {\it et al.} reported new evidence supporting the existence of the hypothetical $X(17)$ boson in the decay of the excited ${^4\text{He}^*}$ state to its ground state \cite{Krasznahorkay:2019lyl}. The $^4\text{He}^*(J^P=0^-) \to {^4\text{He}}(0^+)$ transition implies that $X(17)$ can be either a $J^P=1^+$ boson (orbital angular momentum $L=1$) or a $J^P=0^-$ boson ($L=0$). 
Subsequently, in 2021, they bombarded $^3\text{H}$ with protons of energies 510, 610, and 900 keV to populate the first and second excited states of $^4\text{He}^*$ \cite{Krasznahorkay:2021joi}, and once more observed such anomalous signals of $X(17)$. The derived average values for the mass of $X(17)$ and the decay width ratio are
\begin{align}
    &m_{X}= 16.94 \pm 0.12(\text{stat}) \pm 0.21(\text{sys}) \text{ MeV}, \\
    &\frac{\Gamma\left(^{4}\text{He}^{*} \rightarrow{^{4}\text{He}} X\right)\operatorname{Br}\left(X \rightarrow e^{+} e^{-}\right)}{\Gamma\left(^{4}\text{He}^{*} \rightarrow{^{4}\text{He}} \gamma\right)} =\left(5.1 \pm 0.13 \right)\times 10^{-6}.
\end{align}
In the rest frame of the $^{4}\text{He}^{*} \rightarrow {^{4}\text{He}} X$ decay \footnote{In the laboratory frame, the produced $^4\text{He}^*$ state moves nonrelativistically with a velocity of $0.011c$ for a proton beam energy of 900 keV. Thus, the assumption of the $^4\text{He}^*$ rest frame is appropriate.}, the velocity of $X(17)$ is $v\approx0.59c$.
Imposing the constraint $L \leq 5$ cm on the decay length, one can estimate a lower limit of $(\varepsilon^v_e)^2 + (\varepsilon^a_e)^2 \gtrsim 7.0 \times 10^{-11}$. For a stricter decay length constraint of $L \leq 1$ cm, one obtains $(\varepsilon^v_e)^2 + (\varepsilon^a_e)^2 \gtrsim 3.5 \times 10^{-10}$.
These two lower limits for $L \leq 5$ cm and $L \leq 1$ cm in the $^4\text{He}^*$ case are approximately twice those derived in the $^8\text{Be}^*$ case, respectively. 
For the pure vector or pure axial-vector coupling scenario, imposing the $L \leq 1$ cm constraint requires $|\varepsilon_e^v|$ or $|\varepsilon_e^a|$ to be greater than approximately $1.9 \times 10^{-5}$.

In 2022, Krasznahorkay {\it et al.} further reported a new anomaly supporting the existence and vector character of the hypothetical $X(17)$ boson \cite{Krasznahorkay:2022pxs}. 
They bombarded $^{11}\text{B}$ with protons of five energies in the range of $1.5 - 2.5$ MeV to study the transition $^{12}\text{C}^*(1^-) \longrightarrow {^{12}\text{C}}(0^+)$, which implies that $X(17)$ can be either a $J^P=1^+$ boson (orbital angular momentum $L=1$), or $J^P=1^-$ boson ($L=0$), or a $J^P=0^+$ boson ($L=1$). 
The average values of the $X(17)$ mass and the decay width ratio are
\begin{align}
    &m_{X}= 17.03 \pm 0.11(\text{stat}) \pm 0.20(\text{sys}) \text{ MeV}, \\
    &\frac{\Gamma\left(^{12}\text{C}^{*} \rightarrow {^{12}\text{C}} X\right)\operatorname{Br}\left(X \rightarrow e^{+} e^{-}\right)}{\Gamma\left(^{12}\text{C}^{*} \rightarrow {^{12}\text{C}} \gamma\right)} =\left(3.6 \pm 0.3 \right)\times 10^{-6}.
\end{align}
In the rest frame of the $^{12}\text{C}^{*} \rightarrow {^{12}\text{C}} X$ decay \footnote{In the laboratory frame, the produced $^{12}\text{C}^*$ state moves nonrelativistically with a velocity of $0.005c$. Thus, the assumption of the $^{12}\text{C}^*$ rest frame is appropriate.}, the velocity of $X(17)$ is $v\approx0.20c$.
Imposing the constraint $L \leq 5$ cm on the decay length, one can then estimate a lower limit of $(\varepsilon^v_e)^2 + (\varepsilon^a_e)^2 \gtrsim 2.0 \times 10^{-11}$. For $L \leq 1$ cm, one obtains $(\varepsilon^v_e)^2 + (\varepsilon^a_e)^2 \gtrsim 9.8 \times 10^{-11}$.
These two lower limits for $L \leq 5$ cm and $L \leq 1$ cm in the $^{12}\text{C}^*$ case are approximately half of those derived in the $^8\text{Be}^*$ case, respectively.
For the pure vector or pure axial-vector coupling scenario, this constraint requires $|\varepsilon_e^v|$ or $|\varepsilon_e^a|$ to be greater than approximately $9.9 \times 10^{-6}$.
In Eq. (F.6) of Ref. \cite{Barducci:2022lqd}, Barducci and Toni derive a constraint of $(\varepsilon^v_e)^2 + (\varepsilon^a_e)^2 \gtrsim 9.8 \times 10^{-13}$ by assuming the prompt and saturated decay of $X(17) \to e^+e^-$ in the Atomki detector. 
Since the selection of decay length $L$ is arbitrary, we refrain from drawing conclusions based on constraints on  $(\varepsilon^v_e)^2 + (\varepsilon^a_e)^2$ derived from decay lengths.

\begin{table}[!htb]
\centering
\begin{tabular}{c|c}
\hline
Decay channels & $J^P$ for $X(17)$ \\
\hline
$\mathrm{^8Be^*}(1^+) \to \mathrm{^8Be}(0^+) + X(17)$ & $0^-,\ 1^+,\ 1^-,\ \dots$ \\
$\mathrm{^4He^*}(0^-) \to \mathrm{^4He}(0^+) + X(17)$ & $0^-,\ 1^+,\ \dots$ \\
$\mathrm{^{12}C^*}(1^-) \to \mathrm{^{12}C}(0^+) + X(17)$ & $0^+,\ 1^+,\ 1^-,\ \dots$ \\
\hline
Intersection & $1^+,\ \dots$ \\
\hline
\end{tabular}
\caption{The allowed $J^P$ quantum numbers of $X(17)$ from three Atomki experiments, where dots denote the suppressed quantum numbers with $L\geq 2$.}  \label{tab:JP}
\end{table}

In Table \ref{tab:JP}, we show the allowed $J^P$ quantum numbers of $X(17)$ from three Atomki experiments. Here, the $J^P$ quantum numbers with orbital angular momentum $L\geq 2$ are suppressed. 
It is found that the $J^P$ intersection of all three Atomki experiments is $1^+$, alternatively the axial-vector coupling of $X(17)$ to electrons is favored.

\subsubsection{Electron's Anomalous Magnetic Momentum}

Since the hypothetical $X(17)$ boson couples to electrons, it will contribute to the electron's anomalous magnetic moment (AMM), defined as $a_e \equiv (g_e - 2)/2$.
In 2022, an improved determination of the electron's AMM was reported \cite{Fan:2022eto},
\begin{equation}
a_{e}^{\exp} = 0.00115965218059\,(13),
\end{equation}
which is 2.2 times more accurate than the 2008 measurement \cite{Hanneke:2008tm}.
For the SM predictions on electron's AMM, there are two recent results obtained using fine-structure constant $\alpha$ extracted from Cesium (Cs) \cite{Parker:2018vye} and Rubidium (Rb) \cite{Morel:2020dww} atomic interferometry experiments in 2018 and in 2020 respectively,
\begin{align}
\left( a_{e}^{\mathrm{SM}} \right)_{\mathrm{Cs}} &= 0.00115965218161\,(23), \\
\left( a_{e}^{\mathrm{SM}} \right)_{\mathrm{Rb}} &= 0.001159652180252\,(95),
\end{align}
which differs from each other by 5.5$\sigma$.
Then we obtain the discrepancies between the experimental measurement and the SM predictions, given by
\begin{align}
\left(\Delta a_{e} \right)_{\mathrm{Cs}} &= a_{e}^{\exp} - \left(a_{e}^{\mathrm{SM}} \right)_{\mathrm{Cs}}= (-102.0 \pm 26.4) \times 10^{-14},\\
\left(\Delta a_{e} \right)_{\mathrm{Rb}} &= a_{e}^{\exp} - \left(a_{e}^{\mathrm{SM}} \right)_{\mathrm{Rb}}= (33.8 \pm 16.1) \times 10^{-14},
\label{eq:ae}
\end{align}
which have $-3.9\sigma$ and $2.1\sigma$ discrepancies for $\mathrm{Cs}$ and $\mathrm{Rb}$ cases, respectively.

Within the ``$V \pm A$'' framework, the $X(17)$ boson contributes to the electron's AMM via the leading one-loop triangle diagram \cite{Leveille:1977rc}, with the contribution given by
\begin{equation}
    a_e^{X} = \frac{\alpha}{3\pi}\left(\frac{m_e}{m_X}\right)^2\left[(\varepsilon_e^v)^2 - 5(\varepsilon_e^a)^2\right].
\end{equation}
To diminish discrepancies between the experimental measurement and the SM predictions, we derive the following constraints
\begin{align}
     \mathrm{Cs}:& \quad (-1.4 \pm 0.4) \times 10^{-6} \lesssim(\varepsilon_e^v)^2 - 5(\varepsilon_e^a)^2 \lesssim 0, \label{eq:aeboundCs} \\
     \mathrm{Rb}:& \quad 0 \lesssim (\varepsilon_e^v)^2 - 5(\varepsilon_e^a)^2 \lesssim (4.8 \pm 2.3) \times 10^{-7},
\label{eq:aebound}
\end{align}
where the uncertainties originate from experimental errors.  

The two constraints in Eqs. \eqref{eq:aeboundCs} and \eqref{eq:aebound} have opposite signs, leading to contradictory conclusions in the pure vector and pure axial-vector models. 
The pure vector $Xee$ coupling would further exacerbate the electron's AMM discrepancy under the $\mathrm{Cs}$ constraint, implying that the pure vector coupling is disfavored. 
In the $\mathrm{Cs}$ case, one can deduce the constraint $|\varepsilon_e^a|_{a_e(\mathrm{Cs)}} \lesssim (5.4 \pm 0.7) \times 10^{-4}$ for the pure axial-vector coupling. 
In contrast, the $\mathrm{Rb}$ constraint disfavors the pure axial-vector coupling model.  
In the $\mathrm{Rb}$ case, one obtains $|\varepsilon_e^v|_{a_e(\mathrm{Rb})} \lesssim (6.9 \pm 1.6) \times 10^{-4}$ for the pure vector coupling.  
Adopting the 2008 measurement and SM prediction from the $\mathrm{Cs}$ experiment of the electron's AMM yields a bound consistent with the result in Ref. \cite{Krasnikov:2019dgh}.

\subsubsection{Parity-violating M{\o}ller Scattering}

\begin{figure}[ht]
\begin{center}
\includegraphics[scale=0.8]{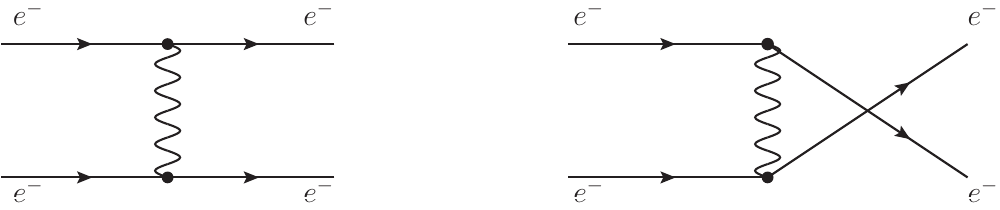}
\caption{The Feynman diagrams for the M{\o}ller Scattering, $e^{-} e^{-} \rightarrow e^{-} e^{-}$, where the propagators can be $\gamma^*$, $Z^0$ or $X(17)$ bosons accordingly.}\label{fig:MullerScattering}
\end{center}
\end{figure}

The parity-violating M{\o}ller scattering can constrain the combination $\varepsilon_e^v \varepsilon_e^a$. 
The left-right scattering asymmetry $A_{PV}$ in fixed-target electron-electron (M{\o}ller) scattering is defined as \cite{Derman:1979zc}
\begin{equation}
    A_{PV} \equiv \frac{d\sigma_R - d\sigma_L}{d\sigma_R + d\sigma_L}, \label{eq:APVdef}
\end{equation}
where $\sigma_{R/L}$ denotes the cross section for incident right/left-handed electrons. 
The Feynman diagrams for M{\o}ller scattering are displayed in Fig. \ref{fig:MullerScattering}, where the propagators can be $\gamma^*$, $Z^0$, or $X(17)$ bosons accordingly. 
At leading order in SM, $A_{PV}$ arises from the interference between the weak and electromagnetic amplitudes, specifically the interference between Feynman diagrams involving $Z^0$ propagation and those with $\gamma^*$ propagation. 
Within the $V \pm A$ model for $X(17)$, the interference between Feynman diagrams involving $X(17)$ propagation and those with $\gamma^*$ propagation will also contribute to $A_{PV}$. 

For the electromagnetic process, the squared amplitudes $|M_{R/L}^{\text{QED}}|^2$ for right/left-handed electrons in the $m_e\to 0$ limit take the compact form
\begin{equation}
        |M_R^{\text{QED}}|^2 = |M_L^{\text{QED}}|^2 = 8 e^2 \left( \frac{s^2}{(s+t)^2} + \frac{s^2}{t^2} + 1\right),\label{eq:MullerQED}
    \end{equation}
where $s$ and $t$ are the Mandelstam variables, and $e$ denotes the electric charge.
In SM, the leading-order contribution to $A_{PV}$ at the amplitude level originates from the difference between the right-handed and left-handed amplitudes for the interference between $Z^0$ and $\gamma^*$ propagated Feynman diagrams. In the $m_e\to 0$ limit, this difference is given by
\begin{equation}
        M_R^{\text{QED}} M_R^{Z *} -  M_L^{\text{QED}} M_L^{Z *} + \text{C.C.} = - \frac{2 e^2 s^3 \left(4s_W^2-1\right)\left(2m_Z^2 + s\right)}{c_W^2 s_W^2 t \left(s+t\right)\left(m_Z^2 -t \right)\left(m_Z^2 +s +t\right)}, \label{eq:MullerQEDZ}
    \end{equation}
where $\text{C.C.}$ stands for the complex conjugated amplitudes, $c_W$ and $s_W$ are the cosine and sine of the Weinberg angle respectively, and $m_Z$ is the mass of the $Z^0$ boson.
Considering the $V \pm A$ model for $X(17)$, its contribution to $A_{PV}$ in the $m_e\to 0$ limit is expressed as
\begin{equation}
        M_R^{\text{QED}} M_R^{X *} -  M_L^{\text{QED}} M_L^{X *} + \text{C.C.} = \frac{32 e^2 \varepsilon_e^v \varepsilon_e^a s^3\left(2 m_X^2 +s \right)}{t \left(s+t\right)\left(m_X^2 -t \right)\left(m_X^2 +s +t\right)},
        \label{eq:MullerQEDX}
    \end{equation}
which is proportional to $\varepsilon_e^v \varepsilon_e^a$.
At the leading order, the left-right scattering asymmetry $A_{PV}$ for fixed-target M{\o}ller scattering is thus given by
\begin{align}
        A_{PV} &= A_{PV}^{Z} + A_{PV}^{X} \nonumber \\
        &= \frac{ M_R^{\text{QED}} M_R^{Z *} -  M_L^{\text{QED}} M_L^{Z *} + \text{C.C.} }{ |M_R^{\text{QED}}|^2 +|M_L^{\text{QED}}|^2 } 
        + \frac{ M_R^{\text{QED}} M_R^{X *} -  M_L^{\text{QED}} M_L^{X *} + \text{C.C.} }{ |M_R^{\text{QED}}|^2 +|M_L^{\text{QED}}|^2 },\label{eq:APVcal}
    \end{align}
where the first term corresponds to the SM contribution, and the second term arises from $X(17)$ contribution.

In 2005, the SLAC E158 experiment performed a precision measurement of the parity-violating asymmetry $A_{PV}$ in fixed-target M{\o}ller scattering \cite{SLACE158:2005uay}, yielding
\begin{equation}
        A_{PV}^{\text{exp.}} = \left(-131 \pm 14 \left(\text{stat.}\right) \pm 10 \left(\text{syst.}\right)\right) \times 10^{-9}, \label{eq:APVexp}
\end{equation}
which enabled the determination of the running weak mixing angle. The measurement was conducted with average values of the kinematic variables $-t = Q^2 = 0.026$ GeV$^2$ and $Q^2/s \simeq 0.6$ \cite{SLACE158:2005uay}.
Using the values $t=-0.026$ GeV$^2$, $\sqrt{s}= 0.208$ GeV, and $m_Z = 91.188$ GeV, we obtain the SM prediction
\begin{equation}
        A_{PV}^{Z} = -122 \times 10^{-9},
    \end{equation}
which is consistent with the experimental measurement $ A_{PV}^{\text{exp.}}$ within the quoted uncertainties. 
To account for the difference between $ A_{PV}^{\text{exp.}}$ and $ A_{PV}^{Z}$, we use $A_{PV}^{X}$ to derive a constraint on $\varepsilon_e^v \varepsilon_e^a$,\footnote{This central value differs by approximately two orders of magnitude from the estimate $\varepsilon_e^v \varepsilon_e^a \lesssim 1.1 \times 10^{-7}$ given in Eq.~(F.3) of Ref.~\cite{Barducci:2022lqd}.}
\begin{equation}
        \varepsilon_e^v \varepsilon_e^a \lesssim \left(2.6\pm 5.0\right) \times 10^{-9}, \label{eq:APVbound}
\end{equation}
where the uncertainties originate from the experimental errors in Eq. \eqref{eq:APVexp}. 
The bound obtained from parity-violating M{\o}ller scattering imposes stringent constraints on the $Xee$ couplings in the $V \pm A$ model.

We could combine the results of several independent experiments into a single constraint on the model parameters by means of a $\chi^2$ analysis. 
In this study the model has two free parameters, the vector and axial-vector couplings $\varepsilon_e^v$ and $\varepsilon_e^a$. 
Each experiment measures an observable $Q_i$ that is a known function of these couplings and reports a central value $\bar{Q}_i$ with an uncertainty $\sigma_i$. 
The $\chi^2$ function combines all measurements as
\begin{equation}
\chi^2(\varepsilon_e^v,\varepsilon_e^a)
= \sum_i \left(\frac{Q_i(\varepsilon_e^v,\varepsilon_e^a)-\bar{Q}_i}{\sigma_i}\right)^2,
\end{equation}
where $Q_i(\varepsilon_e^v,\varepsilon_e^a)$ is the value of the $i$-th observable implied by the couplings. 
In our case the observables are $(\varepsilon_e^v)^2-5(\varepsilon_e^a)^2$, measured by the Cs and Rb atomic interferometry experiments, and the parity-violating product $\varepsilon_e^v\varepsilon_e^a$, measured by M{\o}ller scattering. 
The best-fit point minimizes $\chi^2$, and the confidence regions follow from the contours of
\begin{equation}
\Delta\chi^2 = \chi^2(\varepsilon_e^v,\varepsilon_e^a)-\chi^2_{\min}.
\end{equation}
With two free parameters, the $68.3\%$ ($1\sigma$) and $95.4\%$ ($2\sigma$) regions correspond to $\Delta\chi^2=2.30$ and $\Delta\chi^2=6.18$, respectively. 
For a measurement that reports only an upper limit, the corresponding term is replaced by the one-sided form
\begin{equation}
\chi^2_{\mathrm{upper}}=
\begin{cases}
0 & Q\le Q_{\mathrm{lim}},\\[4pt]
\left(\dfrac{Q-Q_{\mathrm{lim}}}{\sigma}\right)^2 & Q>Q_{\mathrm{lim}},
\end{cases}
\end{equation}
which vanishes below the limit and grows quadratically above it.

\begin{figure}[!htb]
    \centering
    \includegraphics[width=0.8\linewidth]{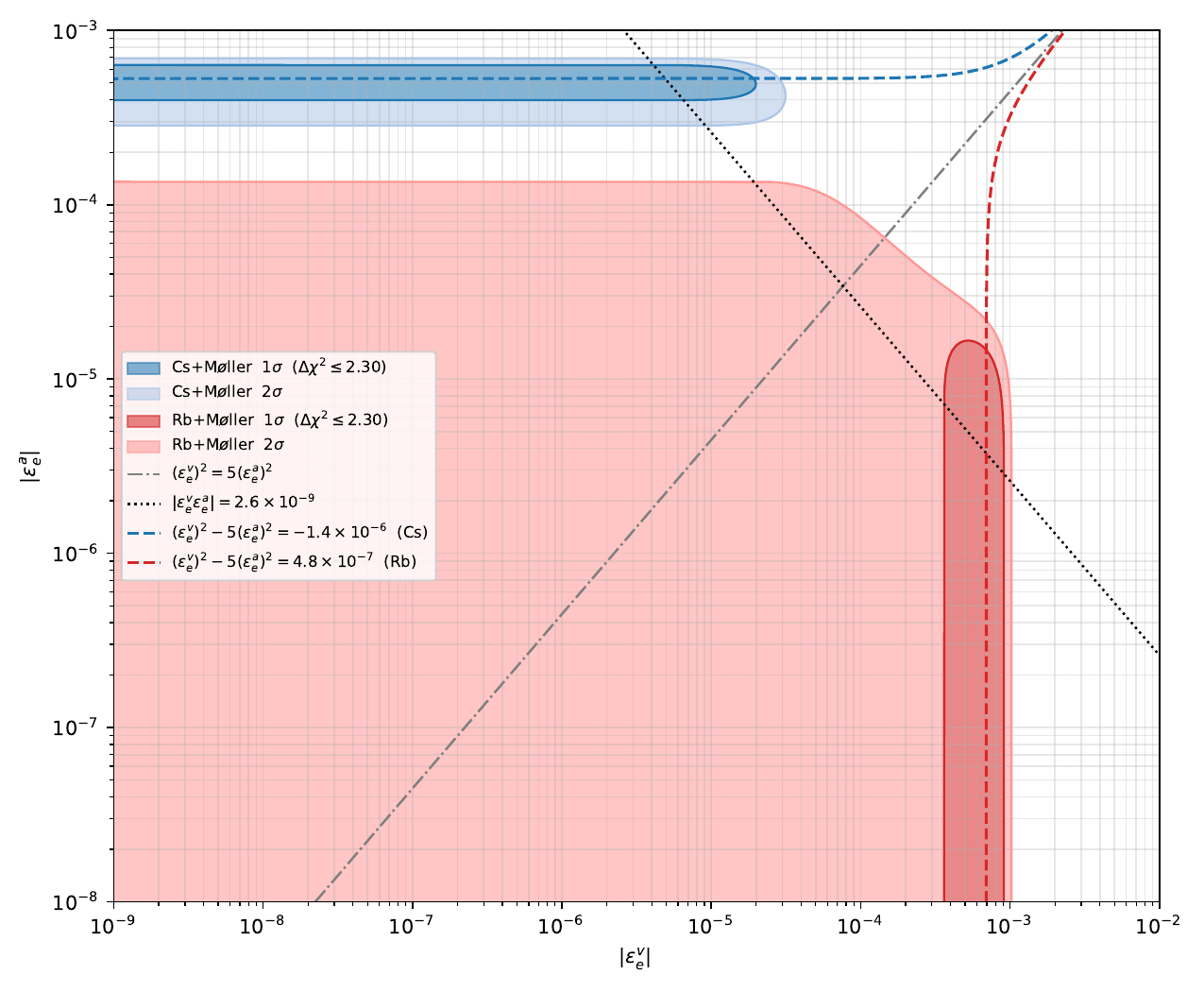}
    \caption{Constraints on $|\varepsilon_e^v|$ and $|\varepsilon_e^a|$ from a joint $\chi^2$ analysis of the electron's AMM measurements (Cs and Rb) and parity-violating M{\o}ller scattering. The blue (red) shaded regions show the $68\%$ and $95\%$ confidence contours of the Cs (Rb) + M{\o}ller combinations.}
    \label{fig:AMMuller}
\end{figure}

We perform a combined $\chi^2$ analysis of the electron's AMM measurements from the Cs and Rb atomic interferometry experiments together with the parity-violating M{\o}ller scattering result. 
The resulting constraints in the $(|\varepsilon_e^v|,|\varepsilon_e^a|)$ plane are presented in Fig. \ref{fig:AMMuller}. 
The two AMM measurements constrain the same combination $Q=(\varepsilon_e^v)^2-5(\varepsilon_e^a)^2$ with opposite signs: the Cs result requires $Q<0$, favouring a predominantly axial-vector coupling, whereas the Rb result requires $Q>0$, favouring a predominantly vector coupling. 
This tension yields two disjoint allowed regions, 
\begin{align}
    a_e(\text{Cs}) + A_{PV}&: |\varepsilon_e^a|\approx5.3\times10^{-4}\gg|\varepsilon_e^v|, \\
    a_e(\text{Rb}) + A_{PV}&: |\varepsilon_e^v|\approx6.9\times10^{-4}\gg|\varepsilon_e^a|. \label{eq:AMMAPV}
\end{align}
In Fig. \ref{fig:AMMuller}, the $95\%$ confidence region of the Rb combination is notably larger because the corresponding contour satisfies $\Delta\chi^2=6.18$, i.e. a $2.49\sigma$ deviation, so at this level the sign of $Q$ is no longer determined and the allowed region spills across $Q=0$ into the axial-vector side. 
Note that the Atomki experiments favor the axial-vector coupling of $X(17)$ to electrons.

\subsubsection{Beam Dump Experiments}

\begin{figure}[ht]
\begin{center}
\includegraphics[scale=0.8]{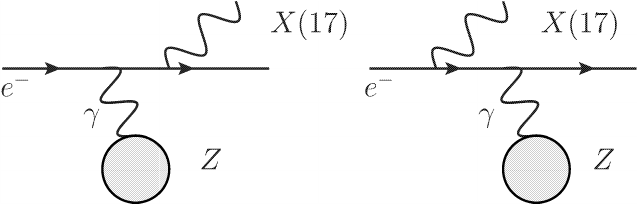}
\caption{The Feynman diagrams for the $e^- Z \to e^- Z X(17)$ process, where the circle represents the electromagnetic interaction with nucleus with the atomic number $Z$.}\label{fig:eZ2eZX}
\end{center}
\end{figure}

In beam dump experiments, the $X(17)$ boson is produced via initial/final-state radiation from a single electron impinging on a fixed target with atomic number $Z$, a process known as the bremsstrahlung reaction $e^- Z \to e^- Z X(17)$, as shown in Fig. \ref{fig:eZ2eZX}. 
When considering the ``$V \pm A$'' interaction between $X(17)$ and electrons in this process, the squared amplitudes are proportional to the coupling parameters solely through the combination $\left(\varepsilon_e^v \right)^2 + \left(\varepsilon_e^a \right)^2$, up to an error of $\mathcal{O}\left(\frac{m_e^2}{m_X^2}\right) \approx 0.1\%$.
More specifically, the amplitude related to the electron chain $e^-(p_1) \to e^-(p_2) + X(p_3) + \gamma^*(p_4)$ is
\begin{align}
    i M &= \frac{\varepsilon^*(p_3)}{p_4^2} \bar{u}(p_2) \left(ie \gamma_\mu\left(\varepsilon_e^v - \varepsilon_e^a \gamma_5 \right) \right) \left(\gamma \cdot \left(p_2 + p_3\right) + m_e\right) \left(ie \gamma_\nu \right) u(p_1) \nonumber \\
    & + \frac{\varepsilon^*(p_3)}{p_4^2} \bar{u}(p_2) \left(ie \gamma_\nu \right) \left(\gamma \cdot \left(p_2 + p_4\right) + m_e\right) \left(ie \gamma_\mu\left(\varepsilon_e^v - \varepsilon_e^a \gamma_5 \right) \right) u(p_1).
\end{align}
Then, in the Dirac trace of the closed electron loop in the squared amplitudes, the term $(\varepsilon_e^v - \varepsilon_e^a \gamma_5)$ must commute an odd number of times across $(\gamma^\alpha + m_e)$ or $\gamma^\alpha$ matrices to contract with its complex conjugate term $(\varepsilon_e^v + \varepsilon_e^a \gamma_5)$. Thus, in the squared amplitudes, the $m_e \to 0$ limit yields two contributions: the combined term $\left(\varepsilon_e^v \right)^2 + \left(\varepsilon_e^a \right)^2$ and the interference term $\varepsilon_e^v\varepsilon_e^a$. However, terms proportional to $\varepsilon_e^v\varepsilon_e^a$ always contain a single $\gamma^5$ in the Dirac trace and therefore vanish upon summation. 
So one can derive the constraints on $\left(\varepsilon_e^v \right)^2 + \left(\varepsilon_e^a \right)^2$ within the $V \pm A$ model from the beam dump experiments which originally constrain the pure vector models.

Recently, the NA64 collaboration has conducted multiple searches for the $X(17)$ boson and set constraints on the $Xee$ coupling parameter, i.e. $\sqrt{\left(\varepsilon_e^v \right)^2 + \left(\varepsilon_e^a \right)^2}$ under the ``$V \pm A$'' assumption up to an error of $\mathcal{O}\left(\frac{m_e^2}{m_X^2}\right)$.
In 2018, the NA64 collaboration first excluded the region $1.3 \times 10^{-4} < \sqrt{\left(\varepsilon_e^v \right)^2 + \left(\varepsilon_e^a \right)^2} < 4.2 \times 10^{-4}$ at the 90\% C.L. \cite{Banerjee:2018vgk}. 
Subsequently, in 2019, a combined analysis of the data samples collected in 2017 and 2018 allowed them to extend the exclusion region to $1.2 \times 10^{-4} < \sqrt{\left(\varepsilon_e^v \right)^2 + \left(\varepsilon_e^a \right)^2} < 6.8 \times 10^{-4}$ at the 90\% C.L. \cite{Banerjee:2019hmi}.\footnote{In Eq.~(F.5) of Ref.~\cite{Barducci:2022lqd}, it is claimed that the NA64 experiment yields the constraint
$\sqrt{\left(\varepsilon_e^v \right)^2 + \left(\varepsilon_e^a \right)^2} \gtrsim 1.2 \times 10^{-4}$,
which is inconsistent with the original NA64 results \cite{Banerjee:2018vgk,Banerjee:2019hmi}.
Our estimate for the upper exclusion limit from the NA64 experiment is consistent with that in Ref.~\cite{Fieg:2026zkg}.
}
In 2020, the NA64 collaboration presented an independent analysis \cite{NA64:2020xxh} that confirmed the 2019 result reported in Ref. \cite{Banerjee:2019hmi}.
In 2021, assuming $X(17)$ is a pseudoscalar boson, they excluded the range $\left( 2.1 \sim 3.2 \right) \times 10^{-4}$ \cite{NA64:2021aiq}.

Additional constraints exist from other beam dump experiments, E137, E141, and E774 experiments, that utilize the bremsstrahlung reaction to search for low-mass dark bosons. 
The data could be extracted from the Fig. 1 of Ref. \cite{Bjorken:2009mm} using WebPlotDigitizer \cite{WebPlotDigitizer}.
In the mass region around 17 MeV, the combined results of the E137 and E141 experiments have excluded the region $8 \times 10^{-8} \lesssim \sqrt{\left(\varepsilon_e^v \right)^2 + \left(\varepsilon_e^a \right)^2} \lesssim 3 \times 10^{-4}$ \cite{Bjorken:2009mm}.\footnote{In Eq.~(F.4) of Ref.~\cite{Barducci:2022lqd}, the lower exclusion limit is reported as $3.6 \times 10^{-8}$, which differs from our result by a factor of roughly two. This discrepancy arises because they extract the E141 experiment data from Figs.~2 and~3 of Ref.~\cite{Andreas:2012mt}, which differ slightly from Fig.~1 of Ref.~\cite{Bjorken:2009mm} we adopt. 
}

In summary, at the 90\% C.L. and up to an error of $\mathcal{O}\left(\frac{m_e^2}{m_X^2}\right)$, beam dump experiments have excluded the range $8 \times 10^{-8} \lesssim \sqrt{\left(\varepsilon_e^v \right)^2 + \left(\varepsilon_e^a \right)^2} \lesssim 6.8 \times 10^{-4}$ for $m_X = 17$ MeV. Thus, we derive a lower limit on the $Xee$ coupling strength,
\begin{equation}
    4.6 \times 10^{-7} \lesssim \left(\varepsilon_e^v \right)^2 + \left(\varepsilon_e^a \right)^2, \label{eq:NA64bound}
\end{equation}
which is obtained solely from the NA64 exclusion limit.
It worth reminding the readers that the parameter space $\left(\varepsilon_e^v \right)^2 + \left(\varepsilon_e^a \right)^2 \lesssim 6.4 \times 10^{-15}$ remains formally allowed; however, we argue that such an extremely small $Xee$ coupling strength cannot account for the multiple Atomki experiments observed with high significance, and thus we disregard this region.
Note that, the constraint in Eq. \eqref{eq:NA64bound} has no error bars since the NA64 experiment only obtains the exclusion limits at the 90\% C.L.
For the pure vector or pure axial-vector scenario, the same constraint applies to $|\varepsilon_e^v|$ or $|\varepsilon_e^a|$, which requires these parameters to be greater than approximately $6.8 \times 10^{-4}$.

\subsubsection{KLOE-2 Experiment}

\begin{figure}[ht]
\begin{center}
\includegraphics[scale=0.8]{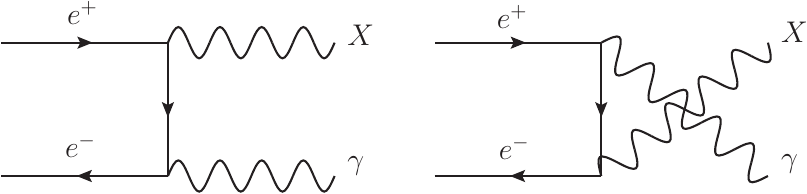}
\caption{The Feynman diagrams of the $X(17)$ boson production in association with a photon in $e^+e^-$ collisions.}\label{fig:ee2xa}
\end{center}
\end{figure}

The KLOE-2 collaboration searched for a low-mass vector boson $U$ in the process $e^+e^-\to U \gamma$ with $U\to e^+e^-$ using the $\mathrm{KLOE}_{(3)}$ dataset at the DA$\Phi$NE $e^+e^-$ collider running at the center-of-mass energy $\sqrt{s} = 1.0195$ GeV  \cite{Anastasi:2015qla}.
The Feynman diagrams for the production of the $X(17)$ boson in association with a photon in $e^+e^-$ collisions are presented in Fig. \ref{fig:ee2xa}. 
In the KLOE-2 experiment, the center-of-mass energy $\sqrt{s}$ is much larger than the masses $m_X$ and $m_e$; additionally, the mass of $X(17)$ is much larger than that of the electron, $(\frac{m_e}{m_X})^2 \approx 0.1\%$. 
Under the ``$V\pm A$'' model, we can thus derive a compact expression for the total cross section,
\begin{equation}
    \sigma = \frac{4\pi\alpha^2}{s}\left( (\varepsilon^v_e)^2 + (\varepsilon^a_e)^2 \right)\left( \log\left[ \frac{s}{m_e^2} \right] - 1 \right) 
    + \mathcal{O}\left( \frac{m_e^2}{m_X^2}, \frac{m_X^2}{s}, \frac{m_e^2}{s} \right), \label{eq:xs}
\end{equation}
which decreases as $\sqrt{s}$ increases. Numerically, the approximated contribution in Eq. \eqref{eq:xs} is indistinguishable from the exact result for $\sqrt{s} = 1.0195$ GeV in the KLOE-2 experiment. It is worth noting that the approximated cross section in Eq. (\ref{eq:xs}) are simply the total cross section of the $e^+e^-\to \gamma \gamma$ process multiplied by the factor $(\varepsilon^v_e)^2 + (\varepsilon^a_e)^2$.

Although the KLOE-2 set an upper bound on the $Xee$ coupling for a low-mass vector boson $U$ at the 90\% C.L. \cite{Anastasi:2015qla}, we can extract the constraints on $(\varepsilon^v_e)^2 + (\varepsilon^a_e)^2$ for $X(17)$ under the $V \pm A$ model by the simple replacement of $|\varepsilon_e^v| \to \sqrt{(\varepsilon^v_e)^2 + (\varepsilon^a_e)^2}$, as qualified explicitly in Eq. \eqref{eq:xs} above. 
Up to an error of $\mathcal{O}\left(\frac{m_e^2}{m_X^2}\right)$, an upper bound can be extracted from Fig. 7 in Ref. \cite{Anastasi:2015qla} using WebPlotDigitizer \cite{WebPlotDigitizer},
\footnote{In Eq.~(F.2) of Ref.~\cite{Barducci:2022lqd}, it is stated that the KLOE experiment yields the constraint
$\sqrt{(\varepsilon_e^v)^2+(\varepsilon_e^a)^2} \lesssim 2.0 \times 10^{-3}$
(or equivalently $(\varepsilon_e^v)^2+(\varepsilon_e^a)^2 \lesssim 4.0 \times 10^{-6}$),
which is consistent with our result and that reported in Ref.~\cite{Fieg:2026zkg}.} 
\begin{equation}
    \left(\varepsilon_e^v \right)^2 + \left(\varepsilon_e^a \right)^2 \lesssim 4 \times 10^{-6}. \label{eq:KLOE2bound}
\end{equation} 
For the pure vector or pure axial-vector scenario, the same upper bound applies to $|\varepsilon_e^v|$ or $|\varepsilon_e^a|$, which is approximately $2 \times 10^{-3}$. 
We could combine the constraints on $\left(\varepsilon_e^v \right)^2 + \left(\varepsilon_e^a \right)^2$ extracted from the KLEO-2 experiment and the NA64 experiment,
\begin{equation}
    4.6 \times 10^{-7} \lesssim \left(\varepsilon_e^v \right)^2 + \left(\varepsilon_e^a \right)^2 \lesssim 4 \times 10^{-6}, \label{eq:NA64KLOE2bound}
\end{equation}
which has no uncertainties since both are obtained at 90\% C.L. in each experiments.

\subsubsection{PADME experiment}

Very recently, the PADME experiment at the Frascati DA$\Phi$NE LINAC searches for $X(17)$ using a positron beam incident on a fixed target \cite{PADME:2025dla}. 
The analysis relies on the measurement of the cross section for the production of events with a two-body final state,  $e^+e^-$ or $\gamma\gamma$ from $e^+e^-$ annihilation processes over a beam energy range of 262-296 MeV, corresponding to center-of-mass energies of $16.4 < \sqrt{s} < 17.4$ MeV. 
This measurement was conducted during a three-month data-taking period in late 2022, covering the relevant mass region indicated by the multiple Atomki experiments.
The most significant deviation was observed at $\sqrt{s} \simeq 16.90$ MeV, with a local significance of $2.5 \, \sigma$ and a global significance of $(1.77 \pm 0.15)\, \sigma$. 
Assuming pure vector coupling, a coupling strength $|\varepsilon_e^v| \approx 1.85 \times 10^{-3}$ was established at the 90\% C.L. 
We remind the reader that this vector coupling strength faces a stringent challenge from the $a_e(Rb)$ constraint $|\varepsilon_e^v|_{a_e(\mathrm{Rb})} \lesssim (6.9 \pm 1.6) \times 10^{-4}$ in the pure vector model.

For the annihilation process $e^+e^- \to X(17)$, the total cross section takes the following simple form,
\begin{equation}
\sigma = \frac{2 \pi^2 \alpha}{s \sqrt{s (s - 4m_e^2)}} 
\left( s \left( (\varepsilon_e ^a)^2 + (\varepsilon_e^v)^2 \right) + 2m_e^2 \left( (\varepsilon_e^v)^2 - \left(3 - \frac{s}{m_X^2}\right)(\varepsilon_e ^a)^2 \right) \right) 
\delta\left(1 - \frac{m_X}{\sqrt{s}}\right),
\end{equation}
where the $\delta$ function arises from the one-particle phase space. 
In the limit $m_e^2 \ll s$, the cross section simplifies to
\begin{equation}
\sigma = \frac{2 \pi^2 \alpha}{s} \left( (\varepsilon_e ^a)^2 + (\varepsilon_e^v)^2 \right) \delta\left(1 - \frac{m_X}{\sqrt{s}}\right) + \mathcal{O}\left( \frac{m_e^2}{s} \right),
\label{eq:ee2X}
\end{equation}
which is proportional to $(\varepsilon_e ^a)^2 + (\varepsilon_e^v)^2$.

\begin{figure}[ht]
\begin{center}
\includegraphics[scale=0.8]{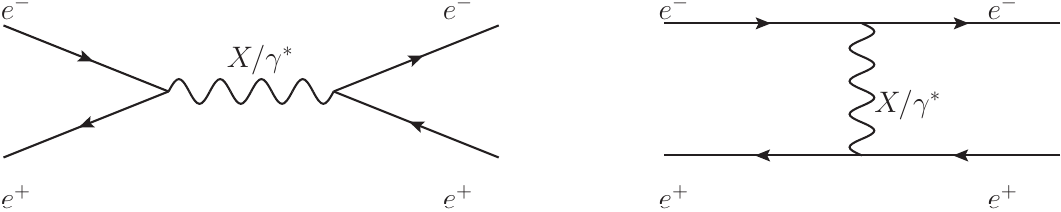}
\caption{The Feynman diagrams for the $e^{+} e^{-} \rightarrow X(17)/\gamma^* \rightarrow e^{+} e^{-}$ processes.}\label{fig:ee2ee}
\end{center}
\end{figure}

For the Bhabha scattering, Feynman diagrams for the signal and background processes, namely $e^{+} e^{-} \rightarrow X(17) \rightarrow e^{+} e^{-}$ and $e^{+} e^{-} \rightarrow \gamma^* \rightarrow e^{+} e^{-}$ are shown in Fig. \ref{fig:ee2ee}.
Under the approximation $m_e^2 \ll s$, the total cross section for the signal process is
\begin{align}
    \sigma &= \frac{4\pi \alpha^2}{3m_X^2 s(s + m_X^2)(s - m_X^2)^2} \nonumber \\
    &\left( s \left( \left( (\varepsilon^v_e)^4 + (\varepsilon^a_e)^4  \right) \left( 3s^3 + 4m_X^2s^2 + m_X^4s - 6 m_X^6\right) \right.\right. \nonumber\\ 
    &\left.\left.\hspace{1cm} + 2(\varepsilon^v_e)^2(\varepsilon^a_e)^2 \left( 3s^3 + 22m_X^2s^2 - 5 m_X^4s - 18 m_X^6 \right) \right)  \right. \nonumber \\ 
    &\left. + 6\left( (\varepsilon^v_e)^4 + (\varepsilon^a_e)^4 + 6(\varepsilon^v_e)^2(\varepsilon^a_e)^2  \right) m_X^2(m_X^2 - s)(m_X^2 + s)^2 \log\left( \frac{m_X^2 + s}{m_X^2} \right) \right) \nonumber \\
    & + \mathcal{O}\left(  \frac{m_e^2}{s} \right),
    \label{eq:xsee2X2ee}
\end{align}
which is a highly reliable approximation.
In the region of $\sqrt{s} \sim m_X$, the total cross section has further simple form
 \begin{equation}
     \sigma(\sqrt{s} \sim m_X) \approx \frac{8\pi \alpha^2 s^2}{3 (s + m_X^2)(s - m_X^2)^2} \left( (\varepsilon^v_e)^2 + (\varepsilon^a_e)^2  \right)^2,
     \label{eq:xsee2X2eeapp}
 \end{equation}
 which is proportional to the squared $(\varepsilon^v_e)^2 + (\varepsilon^a_e)^2$ only. 
 This is understandable because the cross section of $e^+e^- \to X(17)$ in Eq. \eqref{eq:ee2X} and the decay width of $X(17) \to e^+e^-$ in Eq. \eqref{eq:decaywidth_approx} are both proportional to $(\varepsilon^v_e)^2 + (\varepsilon^a_e)^2$.
With the qualification of this proper approximation of Eq. \eqref{eq:xsee2X2eeapp}, one can derive the constraint on $(\varepsilon^v_e)^2 + (\varepsilon^a_e)^2$ from the PADME experiment by the simple replacement of $|\varepsilon_e^v| \to \sqrt{(\varepsilon^v_e)^2 + (\varepsilon^a_e)^2}$ in the $\sqrt{s} \sim m_X$ region, 
\begin{equation}
    \left(\varepsilon_e^v \right)^2 + \left(\varepsilon_e^a \right)^2 \approx 3.4 \times 10^{-6}. \label{eq:PADMEbound}
\end{equation}
However, we refrain from drawing a conclusion based on Eq. \eqref{eq:PADMEbound}. 
Because first the analysis of the PADME experiment relies on both Bhabha scattering and the two-photon final state, whereas Eq. \eqref{eq:xsee2X2eeapp} only validates the extraction of constraints on $(\varepsilon^v_e)^2 + (\varepsilon^a_e)^2$ from Bhabha scattering;\footnote{The two-photon final state process $e^+e^- \to X(17) \to \gamma\gamma$ involves the $X\gamma\gamma$ interaction of $X(17)$ with photons, which is beyond the scope of the $V \pm A$ interaction of the $Xee$ coupling.} second, the constraint from the PADME experiment is only a rough estimate with a statistical significance of approximately $2\sigma$.

\subsection{Parameter Space for $X(17)$}
\label{subsec:survivalee}

\begin{table}[!htb]
    \centering
    \begin{tabular}{c|c|c|c}
    \hline
            &  $V$ &  $A$ & $V \pm A$  \\  \hline \hline
    $a_e(\mathrm{Cs})$  &     $disfavor$      &   $ |\varepsilon_e^a| \mathbf{ \lesssim (5.4 \pm 0.7) \times 10^{-4}}$    &  $\mathbf{(-1.4 \pm 0.4) \times 10^{-6} \lesssim} (\varepsilon_e^v)^2 - 5(\varepsilon_e^a)^2 \lesssim \mathbf{0}$ \\  \hline
    $a_e(\mathrm{Rb})$  &     $|\varepsilon_e^v| \mathbf{\lesssim (6.9 \pm 1.6) \times 10^{-4}}$      &   $disfavor$    &  $\mathbf{0} \lesssim (\varepsilon_e^v)^2 - 5(\varepsilon_e^a)^2 \mathbf{\lesssim (4.8 \pm 2.3) \times 10^{-7}}$ \\  \hline
   $A_{PV}$ & / & / & $\varepsilon_e^v \varepsilon_e^a \mathbf{ \lesssim \left(2.6\pm 5.0\right) \times 10^{-9}}$ \\ \hline \hline
  NA64 &   $6.8 \times 10^{-4} \lesssim |\varepsilon_e^v| $       &   $6.8 \times 10^{-4} \lesssim |\varepsilon_e^a| $    &    $ 4.6 \times 10^{-7} \lesssim \left(\varepsilon_e^v \right)^2 + \left(\varepsilon_e^a \right)^2$     \\  \hline 
   KLEO-2 & $|\varepsilon_e^v| \lesssim 2 \times 10^{-3} $ & $ |\varepsilon_e^a| \lesssim 2 \times 10^{-3} $ & $\left(\varepsilon_e^v \right)^2 + \left(\varepsilon_e^a \right)^2 \lesssim 4 \times 10^{-6}$ \\  \hline
   PADME & $|\varepsilon_e^v| \approx 1.85 \times 10^{-3}$ & $|\varepsilon_e^a| \approx 1.85 \times 10^{-3}$  & $\left(\varepsilon_e^v \right)^2 + \left(\varepsilon_e^a \right)^2 \approx 3.4 \times 10^{-6}$ \\     
   \hline
    \end{tabular}
    \caption{The constraints on the pure vector ($V$), pure axial-vector ($A$), and $V \pm A$ models derived from experiments. Quantities with uncertainties are in bold, while others of NA64, KLEO-2 and PADME experiments are obtained at 90\% C.L.}
    \label{tab:constraints}
\end{table}

\begin{figure}[!htb]
    \centering
    \includegraphics[width=0.8\linewidth]{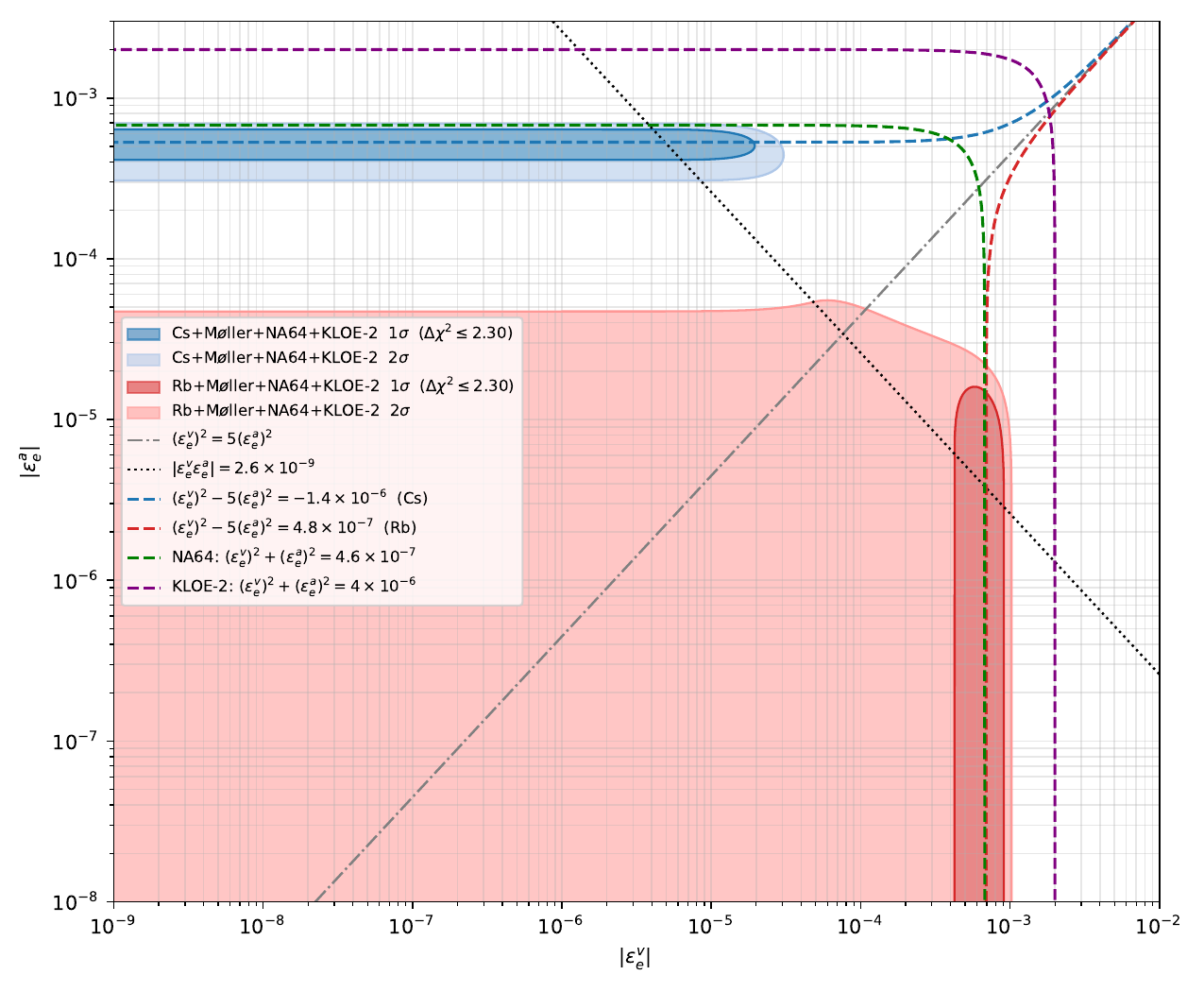}
    \caption{Constraints on $|\varepsilon_e^v|$ and $|\varepsilon_e^a|$ from a joint $\chi^2$ analysis of the electron's AMM measurements (Cs and Rb), parity-violating M{\o}ller scattering, NA64 experiment and KLEO-2 experiment.}
    \label{fig:AMMullerNA64KLEO2}
\end{figure}

We summarize the constraints obtained from previous subsections and discuss the survival parameter spaces for the $X(17)$ hypothesis.
The explicit constraints on the coupling strengths of the pure vector, pure axial-vector, and $V \pm A$ models are presented in Table \ref{tab:constraints}, where the values with uncertainties of electron's AMM and M{\o}ller scattering ($A_{PV}$) are in bold.
To incorporate the NA64 and KLOE-2 bounds in Eq. \eqref{eq:NA64KLOE2bound}, which are reported as $90\%$~C.L. limits without quoted uncertainties, we convert them into effective
$1\sigma$ constraints. 
A one-sided $90\%$~C.L. corresponds to $1.28\sigma$, so the NA64 lower bound and the KLOE-2 upper bound are implemented as one-sided $\chi^2$ terms with effective errors, 
\begin{equation}
    \left(4.6 \pm 3.6 \right) \times 10^{-7} \lesssim \left(\varepsilon_e^v \right)^2 + \left(\varepsilon_e^a \right)^2 \lesssim \left(4 \pm 3.1 \right) \times 10^{-6}. \label{eq:NA64KLOE2bounderror}
\end{equation}

We then re-perform the joint $\chi^2$ analysis after taking the results of NA64 and KLOE-2 experiments into consideration. 
For the $a_e(\text{Cs})+A_{PV}$+NA64+KLEO-2 combination, the best fit is axial-vector dominated,
\begin{equation}
    |\varepsilon_e^a|\approx5.3\times10^{-4},~|\varepsilon_e^v|\lesssim2\times10^{-5} ~ (1\sigma,~\chi^2_{\min}=0.24). \label{eq:surv-A}
\end{equation}
For the $a_e(\text{Rb})+A_{PV}$+NA64+KLEO-2 combination the best fit is vector dominated,
\begin{equation}
    |\varepsilon_e^v|\approx6.9\times10^{-4},~|\varepsilon_e^a|\lesssim1.6\times10^{-5} ~ (1\sigma,~\chi^2_{\min}\approx0). \label{eq:surv-V}
\end{equation}
The corresponding $1\sigma$ and $2\sigma$ contours
($\Delta\chi^2=2.30$ and $6.18$ for two parameters) are shown in Fig. \ref{fig:AMMullerNA64KLEO2}.
Comparing with the fit in Fig. \ref{fig:AMMuller} based only on the electron's AMM (Cs or Rb) and M{\o}ller scattering, the inclusion of NA64 and KLOE-2 modifies only mildly the allowed regions. 
The NA64 lower bound cuts away the small radius part of both regions, raising the lower edge of the Cs $1\sigma$ band in $|\varepsilon_e^a|$ from $4.0\times10^{-4}$ to $4.1\times10^{-4}$, and the lower edge of the Rb $1\sigma$ band in $|\varepsilon_e^v|$ from $3.7\times10^{-4}$ to $4.3\times10^{-4}$. 
Only the Cs+M{\o}ller+NA64+CLEO-2 combination acquires a nonzero $\chi^2_{\min}$, rising from $\approx0$ to $0.24$, because its best-fit value lies below the NA64 lower bound, whereas the Rb+M{\o}ller+NA64+CLEO-2 combination is unchanged since its best fit already satisfies it. The KLOE-2 upper bound is too loose to constrain either combination. 
Note again, three Atomki experiments favor the axial-vector $Xee$ coupling, or $1^+$ $X(17)$ hypothesis.

\section{Summary and Prospects}
\label{sec:sum}

The Atomki experiments have proposed the $X(17)$ particle to account for the observed anomalies in the electron-positron angular correlations from the decays of excited $^{8}\text{Be}$, $^{4}\text{He}$, and $^{12}\text{C}$ nuclei.
In this work, we study the constraints on the couplings of $X(17)$ to electrons from Atomki experiments, the electron's AMM $a_e$, the parity-violating M{\o}ller scattering experiment, beam dump experiments, the KLOE-2 experiment, and the PADME experiment. 
A joint $\chi^2$ analysis of the electron's AMM and the parity-violating M{\o}ller scattering result yields two mutually exclusive solutions, because the Cs and Rb AMM constrain the same combination $(\varepsilon_e^v)^2-5(\varepsilon_e^a)^2$ with opposite signs. 
The Cs+M{\o}ller combination favours a dominant axial-vector coupling with $|\varepsilon_e^a|\approx5.3\times10^{-4}\gg|\varepsilon_e^v|$, whereas the Rb+M{\o}ller combination favours a dominant vector coupling with $|\varepsilon_e^v|\approx6.9\times10^{-4}\gg|\varepsilon_e^a|$. 
The subdominant coupling in each case is suppressed by the M{\o}ller bound $\varepsilon_e^v\varepsilon_e^a<2.6\times10^{-9}$.
Incorporating the NA64 and KLOE-2 bounds at $90\%$ C.L. on $(\varepsilon_e^v)^2+(\varepsilon_e^a)^2$ introduces only a mild tension with the Cs+M{\o}ller solution and leaves the Rb+M{\o}ller solution nearly unchanged.

More dedicated experiments are urgently needed to resolve the Atomki anomalies.
The survival vector dominant coupling strength in Eqs. \eqref{eq:AMMAPV} and \eqref{eq:surv-V} is right above the exclusion region of NA64 experiment \cite{Banerjee:2019hmi}. 
The precise measurements on the $X(17)$-induced Lamb shifts and hyperfine structures of atoms might be one of the promising research plans \cite{Lin:2026tub}.
In addition, $e^+e^-$ accelerator experiments searching for $X(17)$ signals provide new insights into the Atomki anomalies. 
The $e^+e^- \to X(17) \to e^+e^-$ and $e^+e^- \to X(17)\gamma \to e^+e^-\gamma$ processes are likely the two most viable channels. Nevertheless, both processes face challenges from massive background events, necessitating high energy resolution. Additionally, for the second process at BESIII, the phase space for the $e^+e^-$ pair (or equivalently, the energetic photon) is extremely limited.
Recently, the potential for the discovery of the $X(17)$ boson at the Super Tau-Charm Factory was discussed in Ref. \cite{M:2025dij}.
These upcomping experiments bring us closer to resolving the Atomki anomalies.

Alternatively, given that nuclear transitions involving the excited states $^8\text{Be}^*$, $^4\text{He}^*$, $^{12}\text{C}^*$, and the Giant Dipole Resonance of $^8\text{Be}^*$ all indicate the same anomalous phenomenon, one may argue that the Atomki anomalies may stem from some poorly understood nuclear effect. 
Could this be related to $\alpha$-clusters? 
Notably, $^8\text{Be}$, $^4\text{He}$, and $^{12}\text{C}$ are all integer multiples of the $\alpha$-particle ($^4\text{He}$ nucleus), a characteristic feature that favors the formation of $\alpha$-cluster structures in these nuclei.
This conjecture could be verified by searching for an unexpected bump in the electron-positron angular correlations during the de-excitation of the excited $^{16}\text{O}^*$ nucleus to its ground state, and by comparing these correlations with those from the de-excitation of excited nuclei lacking $\alpha$-cluster structures. Several suitable excited states of $^{16}\text{O}^*$ lie above 17 MeV, as listed in Table 16.13 of Ref. \cite{Tilley:1993zz}. 
Very recently, unexpectedly abundant $\alpha$ clustering is found in warm and dense nuclear matter from heavy-ion collisions \cite{Wang:2025wim}.

\begin{acknowledgments}

We appreciate very much the anonymous reviewers' valuable suggestions of deriving the constraints from the parity-violating M{\o}ller scattering experiment and from the electron's anomalous magnetic moment in Rubidium interferometry experiment, and conduct a joint $\chi^2$ analysis on available experimental constraints on $Xee$ couplings, which improve our manuscript significantly. 
We thank Prof. Jie Zhao who organized the workshop on $X(17)$ particle held in Fudan University on Aug. 11th, 2025.
This work is supported in part by the National Key Research and Development Program of China under
Contracts No. 2025YFA1613900, by the National Natural Science Foundation of China (NSFC) under
the Grants 12475083, 12475087, 12235008.

\end{acknowledgments}


\bibliography{XDetection}

\end{document}